\documentclass[fleqn,aps,pra,amsmath,amssymb,reprint,showpacs]{revtex4-1}

\usepackage{amsmath}
\usepackage{graphicx}
\usepackage{dcolumn}
\usepackage{bm}
\usepackage{textcomp}
\usepackage{xcolor}

\def\bra#1{\mathinner{\langle{#1}|}}
\def\ket#1{\mathinner{|{#1}\rangle}}

\renewcommand\Re{\operatorname{Re}}
\renewcommand\Im{\operatorname{Im}}

\begin{document}

\title{Study of a four-level system in vee + ladder configuration}
 \author{Vineet Bharti}
 \affiliation{Department of Physics, Indian Institute of
 Science, Bangalore 560012, India}
 \author{Vasant Natarajan}
 \affiliation{Department of Physics, Indian Institute of
 Science, Bangalore 560012, India}
 \email{vasant@physics.iisc.ernet.in}
 \homepage{www.physics.iisc.ernet.in/~vasant}

\begin{abstract}
We present the results of a theoretical study of a four-level atomic system in vee + ladder configuration using a density matrix analysis. The absorption and dispersion profiles are derived for a weak probe field and for varying strengths of the two strong control fields. For specificity, we choose energy levels of $^{87}$Rb, and present results for both stationary atoms and moving atoms in room temperature vapor. An electromagnetically induced absorption (EIA) peak with negative dispersion is observed at zero probe detuning when the control fields have equal strengths, which switches to electromagnetically induced transparency (EIT) with positive dispersion (due to splitting of the EIA peak) when the control fields are unequal. There is significant linewidth narrowing in thermal vapor. 
\end{abstract}


\maketitle

\section{Introduction}
The phenomenon of electromagnetically induced transparency (EIT), where the absorption of a weak probe laser on one transition is reduced using a strong control laser on an auxiliary transition, is well-studied in three-level systems, in all the three configurations namely lambda ($\Lambda$), vee (V), and ladder ($\Xi$) \cite{HAR97,HFI90,FIM05}. But the related phenomenon of electromagnetically induced absorption (EIA)---enhanced absorption of the probe laser---is possible only when there are at least four levels, so that additional control lasers can be used. EIA has been studied---both theoretically and experimentally---in N-type ($\Lambda + {\rm V}$) four-level systems \cite{GWR04,BMW09,CPN12}. The other four-level systems, namely Y-type ($ \Xi + \Xi $), inverted Y-type ($ \Lambda + \Xi $), and tripod-type ($ \Lambda + \Lambda $), only show enhanced EIT.

In this work, we present theoretical analysis of a new kind of four-level system formed by the combination of V and $ \Xi $ three-level systems, which shows EIA resonances. This configuration opens up the possibility of the experimental study of EIA in an important class of atoms called Rydberg atoms, because such atoms have a highly excited atomic state allowing the formation of a $\Xi$-type system. Rydberg atoms have thus been used to study EIT phenomena \cite{PMG10,POF11}, and the present analysis allows the extension to the observation of EIA in such atoms.

For application to a real system, we choose energy levels of $^{87}$Rb, and present results for both stationary atoms and moving atoms in room temperature vapor. An EIA peak with negative dispersion is observed at zero probe detuning when the control fields have equal strengths. It \textbf{switches} to an EIT peak with positive dispersion (due to splitting of the EIA peak) when the control fields are unequal. Thus, the atomic medium can be switched from EIA and super-luminal light propagation to EIT and sub-luminal light propagation.

\section{Theoretical Considerations}
\begin{figure}
\centerline{\resizebox{0.7\columnwidth}{!}{\includegraphics{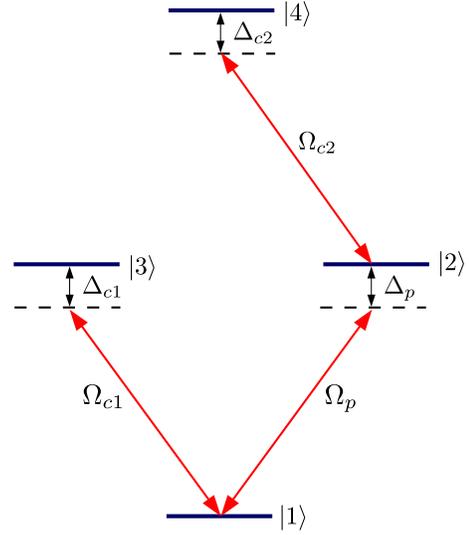}}}
\caption{(Color online) Four-level system formed by the combination of three-level vee and ladder systems.}
\label{levels}
\end{figure}

The four-level system considered in the present work is shown in Fig.\ \ref{levels}. The ground state $\ket{1}$ is coupled with state $\ket{2}$ by a weak probe field. Two strong control fields are present---control 1 between levels $\ket{1}$ and $\ket{3}$, and control 2 between levels $\ket{2}$ and $\ket{4}$. The control fields have detunings $\Delta_{c1}$ and $\Delta_{c2}$, while the frequency of the probe field is scanned with variable detuning $\Delta_{p}$. The intensities in the various beams are given in terms of the respective Rabi frequencies, which are $\Omega_{p}$, $\Omega_{c1}$, and $\Omega_{c2}$.

The total Hamiltonian after carrying out the rotating wave approximation (RWA) is:
\begin{eqnarray}
H = \dfrac{\hbar}{2}\left[\Omega_{p} \ket{1} \bra{2}+\Omega_{c1} \ket{1} \bra{3}+\Omega_{c2} \ket{2} \bra{4}\right]+ {\rm h.c.} \nonumber \\
+\hbar \left[\Delta_{p} \ket{2} \bra{2}+\Delta_{c1} \ket{3} \bra{3}
+(\Delta_{p}+\Delta_{c2}) \ket{4} \bra{4}\right]
\label{Eq.1}
\end{eqnarray}
where h.c.\ is the Hermitian conjugate of the preceding off-diagonal terms. As expected, in the absence of the control 1 (or the control 2) field, the four-level system reduces to a three-level ladder (or vee) system. The time evolution of the system is given by the standard Liouville equation for the density matrix $\rho$. The coupled density matrix equations for this four-level system are given in the appendix.

The observable in our work is the response of atoms to the weak probe field. The susceptibility of probe field is determined by the coherence between levels $\ket{1}$ and  $\ket{2}$---$\rho_{12}$---and is defined as \cite{CJG07}
\begin{equation}
\chi=k\rho_{12} \ \ \text{where} \ \ k= \frac{N|\mu_{12}|^2}{\hbar\varepsilon_{0}\Omega_{p}}
\end{equation}
Here $N$ is the atom number density in the medium and $\mu_{12}$ is the dipole matrix element for the probe transition. The dispersion of the probe field is proportional to $\Re\{\rho_{12}\}$, and its absorption is proportional to $\Im\{\rho_{12}\}$.

The group velocity of the probe field is given by
\begin{equation}
v_{g} = \dfrac{c}{1+\dfrac{1}{2} \Re \chi + \dfrac{\omega_{p}}{2} \dfrac{ \partial \Re \chi }{\partial\omega_{p} } }
\label{Eq.3}
\end{equation}
where $c$ is the velocity of light and $\omega_p$ is the probe frequency. The above equation shows that $v_{g}$ is inversely proportional to the slope of probe dispersion---positive dispersion (EIT in the imaginary part) gives rise to sub-luminal propagation, while negative dispersion (EIA in the imaginary part) gives rise to super-luminal propagation.

The various density-matrix elements are solved following the procedure given in Ref.\ \cite{BHW12}. The steady-state solution for $\rho_{12}$ is given by

\begin{widetext}
\begin{eqnarray}
&\rho_{12}=\displaystyle{-\frac{i\Omega_{p}\rho_{11}}{2\gamma_{12}\beta}+
\frac{i\Omega_{p}|\Omega_{c1}|^2(\rho_{11}-\rho_{33})}{8\gamma_{12}\gamma_{31}\gamma_{32}\beta}
\left(1-\frac{|\Omega_{c2}|^2}{4\gamma_{32}\gamma_{34}\alpha}-
\frac{|\Omega_{c2}|^2}{4\gamma_{14}\gamma_{34}\alpha}\right)}
\label{Eq.4}
\end{eqnarray}
where
\begin{eqnarray*}
&\beta=1+\displaystyle{\frac{|\Omega_{c1}|^2}{4\gamma_{12}\gamma_{32}}+\frac{|\Omega_{c2}|^2}{4\gamma_{12}\gamma_{14}}
-\frac{|\Omega_{c1}|^2|\Omega_{c2}|^2}{16\gamma_{12}\gamma_{34}\alpha}\left[\frac{1}{\gamma_{32}}+\frac{1}{\gamma_{14}}\right]^2}, \alpha=\displaystyle{1+\frac{|\Omega_{c1}|^2}{4\gamma_{14}\gamma_{34}}+\frac{|\Omega_{c2}|^2}{4\gamma_{32}\gamma_{34}}}, \nonumber\\
&\gamma_{12}=\displaystyle{\left(-\frac{\Gamma_{2}}{2}+i\Delta_{p}\right)},
\gamma_{31}=\displaystyle{\left(-\frac{\Gamma_{3}}{2}-i\Delta_{c1}\right)},
\gamma_{14}=\displaystyle{\left(-\frac{\Gamma_{4}}{2}+i(\Delta_{p}+\Delta_{c2})\right)},   \nonumber \\ &\gamma_{32}=\displaystyle{\left(-\frac{\Gamma_{2}+\Gamma_{3}}{2}-i(\Delta_{c1}-\Delta_{p})\right)},
\gamma_{34}=\displaystyle{\left(-\frac{\Gamma_{3}+\Gamma_{4}}{2}+i(\Delta_{p}+\Delta_{c2}-\Delta_{c1})\right)}.
\end{eqnarray*}
\end{widetext}
Also, the  population difference between states $\ket 1$ and $\ket 3$ is given by \cite{MES99}
\begin{equation}
\label{popdiff}
\rho_{11}-\rho_{33}
= \left[ 1 + \dfrac{|\Omega_{c1}|^2}{2\left(\dfrac{\Gamma_{3}^2}{4}+\Delta_{c1}^2\right)} \right]^{-1}
\end{equation}
As expected, the above equation yields the solutions for three-level $\Xi$ and V-systems by considering that only the corresponding control fields are non-zero \cite{KPW05,DPW06}.

For specificity, we have chosen energy levels of the $^{87}$Rb atom---level $ \ket{1} $ is the $F=1$ hyperfine level of the 5S$_{1/2}$ ground state; level $\ket{2} $ is the $ F=2$ hyperfine level of the 5P$_{3/2}$ state; level $\ket{3} $ is the $ F=2$ hyperfine level of the 5P$_{1/2}$ state; and level $\ket{4}$ is the $ F=1$ hyperfine level of the 5D$_{5/2}$ state. The corresponding decay rates are: $\Gamma_1 =0 $ (because it is a ground state), $ \Gamma_2 = 2 \pi \times 6.1 $ MHz,  $\Gamma_3 = 2 \pi \times 5.9$ MHz, and $\Gamma_4 = 2\pi \times 0.68$ MHz. Since the probe field is weak, it ensures that $ \rho_{22}=0$, and hence $\rho_{44}$ is also $0$. The population cycles between $\ket{1}$ and $\ket{3} $ (closed transition) according to Eq.\ \eqref{popdiff}. The energy levels considered here do not allow the formation of the well-known diamond configuration dealt with in references \cite{MFO02} and \cite{PAW15}, because the 5D$_{5/2}$ state does not couple to the 5P$_{1/2}$ intermediate state.

The above analysis is also applicable to Rydberg atoms with minor changes in parameters for level $\ket{4}$. For example, level $\ket{4}$ can be the $\rm 44 \, {D}_{5/2}$ state considered in Ref.\ \cite{MJA07} with a decay rate of $\Gamma_4 \approx 2\pi \times 0.3 $ MHz. 
\section{Results and discussion}

\subsection{Stationary atoms}
\begin{figure}
\includegraphics[width=0.48\textwidth]{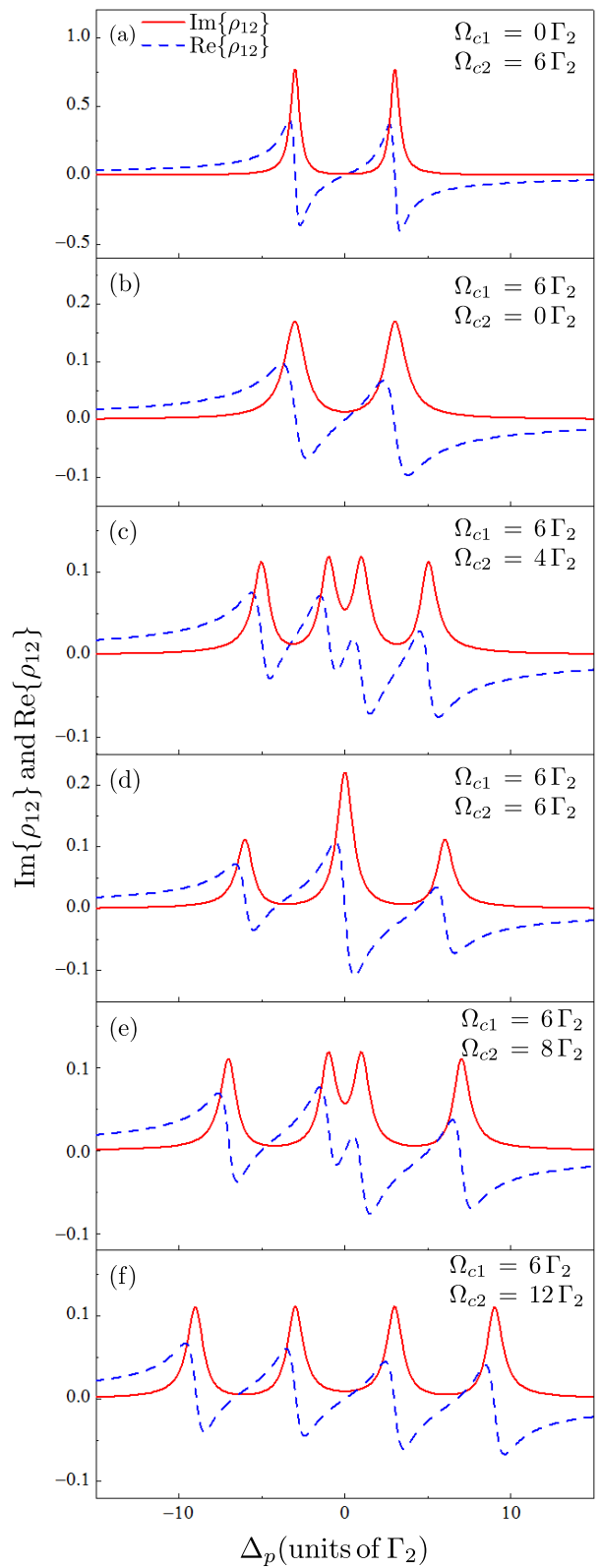}
\caption{(Color online) Calculated probe absorption ($\Im \{ \rho_{12} \}$, solid), and dispersion ($\Re \{ \rho_{12} \} $, dashed) versus probe detuning for (a) $\Omega_{c1} = 0$, $\Omega_{c2} = 6 \, \Gamma_{2}$; (b) $\Omega_{c1} = 6 \, \Gamma_{2}$, $\Omega_{c2} = 0$; (c) $\Omega_{c1} = 6 \, \Gamma_{2}$, $\Omega_{c2} = 4 \, \Gamma_{2}$; (d) $\Omega_{c1} = 6 \, \Gamma_{2}$, $\Omega_{c2} = 6 \, \Gamma_{2}$; (e) $\Omega_{c1} = 6 \, \Gamma_{2}$, $\Omega_{c2} = 8 \, \Gamma_{2}$; (f) $\Omega_{c1} = 6 \, \Gamma_{2}$, $\Omega_{c2} = 12 \, \Gamma_{2}$. }
\label{absdisp}
\end{figure}

We first consider the results for stationary atoms. The imaginary and the real parts of the solution---corresponding to absorption and dispersion, respectively---as a function of probe detuning are shown in Fig.\ \ref{absdisp}. The two control fields are taken to be on resonance, i.e.\ $\Delta_{c1} = \Delta_{c2} = 0 $. Parts (a) and (b) show the results when only one control field is present, i.e.\ corresponding to ladder and V-type three-level systems respectively. As expected, probe absorption shows an EIT dip near line center, because absorption splits into an Autler-Townes doublet at the locations of the dressed states created by the control field, i.e.\ when $\Delta_p = \pm \Omega_c /2$. The corresponding dispersion profile has a positive slope near zero detuning, and hence can be used for \textbf{sub-luminal} light propagation.

Now look at what happens when both control fields are applied. The results for fixed value of $\Omega_{c1} = 6 \, \Gamma_2 $ and varying $\Omega_{c2} $ are shown in part (c)--(f) of the figure. When the two control fields are equal, the medium switches from transmitting to absorbing, and has an EIA peak at line center. The corresponding dispersion profile has negative slope near zero detuning, which can therefore be used for \textbf{super-luminal} light propagation. On either side of this equality condition, the central EIA peak splits into two, and shows a transparency dip at line center. The corresponding dispersion profile has positive slope. Therefore, the medium can be switched from transparency (slow light) to absorption (fast light) by changing the relative strengths of the two control fields.

\subsection{Thermal averaging for moving atoms}

We now consider the results of Doppler averaging for an atomic vapor at room temperature, which is the typical condition for Rb atoms used for an experimental realization. Since the probe and control beams are taken to be co-propagating or counter-propagating, the relevant distribution is the  one-dimensional Maxwell--Boltzmann velocity distribution. For an atom of mass $M$ at a temperature $T$ and moving with a velocity $v$, this is given by
\begin{equation*}
\begin{aligned}
f(v) \, dv &= \sqrt{ \dfrac{M}{2\pi k_BT}} \exp \left( -\dfrac{Mv^2}{2k_BT} \right) dv
\end{aligned}
\end{equation*}
where $k_B$ is the Boltzmann constant. If the velocity is along the beam direction, then the detuning of the field will change by the Doppler shift of $ \pm kv $, where $k$ is the photon wavevector and the sign depends on the relative direction of the beam with respect to the velocity. The thermal averaging over the Maxwell--Boltzmann distribution is done by assuming that the probe field and control field 1 are co-propagating, and control field 2 is counter-propagating.

\begin{figure}
\includegraphics[width=0.42\textwidth]{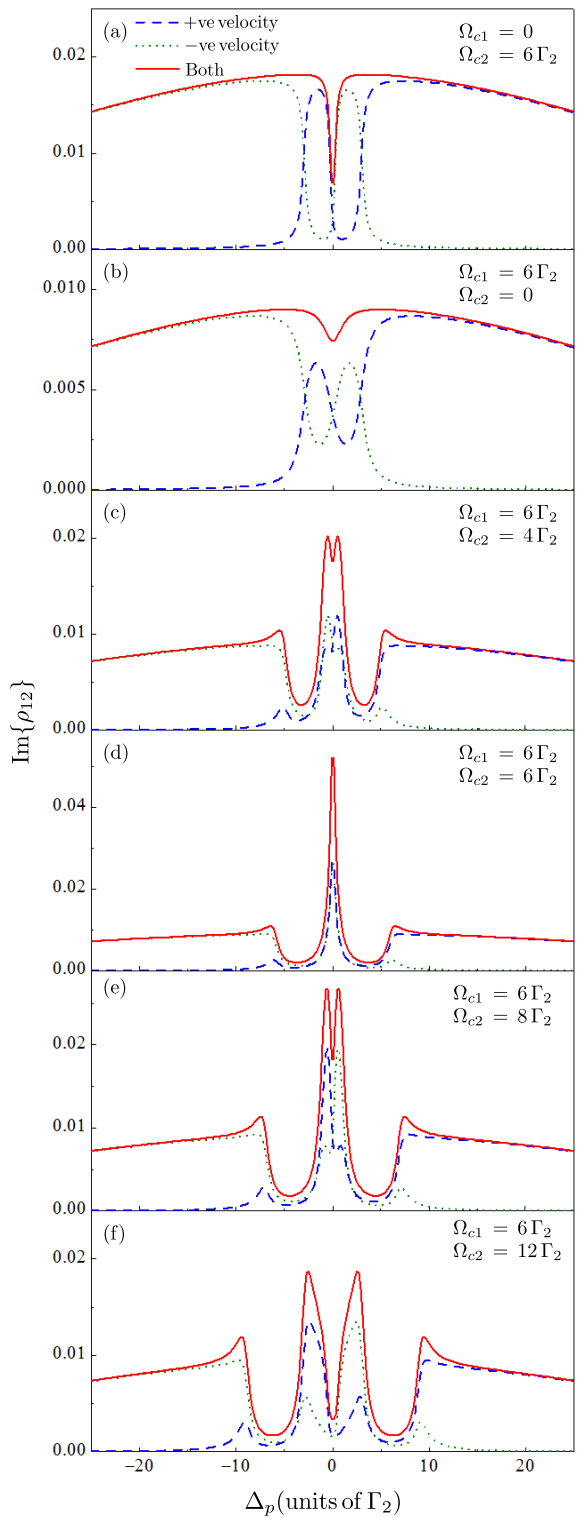}
\caption{(Color online) Effect of Doppler averaging on probe \textbf{absorption} at room temperature. (a) $\Omega_{c1} = 0$, $\Omega_{c2} = 6 \, \Gamma_{2}$; (b) $\Omega_{c1} = 6 \, \Gamma_{2}$, $\Omega_{c2} = 0$; (c) $\Omega_{c1} = 6 \, \Gamma_{2}$, $\Omega_{c2} = 4 \, \Gamma_{2}$; (d) $\Omega_{c1} = 6 \, \Gamma_{2}$, $\Omega_{c2} = 6 \, \Gamma_{2}$; (e) $\Omega_{c1} = 6 \, \Gamma_{2}$, $\Omega_{c2} = 8 \, \Gamma_{2}$; (f) $\Omega_{c1} = 6 \, \Gamma_{2}$, $\Omega_{c2} = 12 \, \Gamma_{2}$. Solid curves are for all atoms, dashed curves are for atoms moving with positive velocities, and dotted curves are for atoms moving with negative velocities.}
\label{Fig.4}
\end{figure}

The absorption profiles after thermal averaging, shown in Fig.\ \ref{Fig.4}, confirm our earlier observation of linewidth reduction in Ref.\ \cite{IFN09}. The main results are the same as that seen for stationary atoms---an EIT dip for a three-level system when one of the control fields has zero strength; an EIA peak when the two control fields are equal; and the EIA peak splitting into two when the control fields have unequal strengths. The linewidth reduction is because moving atoms fill in the transparency window. This can be seen in the figure, where the profile for atoms moving only with positive velocities is shown with a dashed line while the profile for atoms moving only with negative velocities is shown with a dotted line---these profiles are calculated after Doppler averaging for one sign of velocity only.

\begin{figure}
\includegraphics[width=0.42\textwidth]{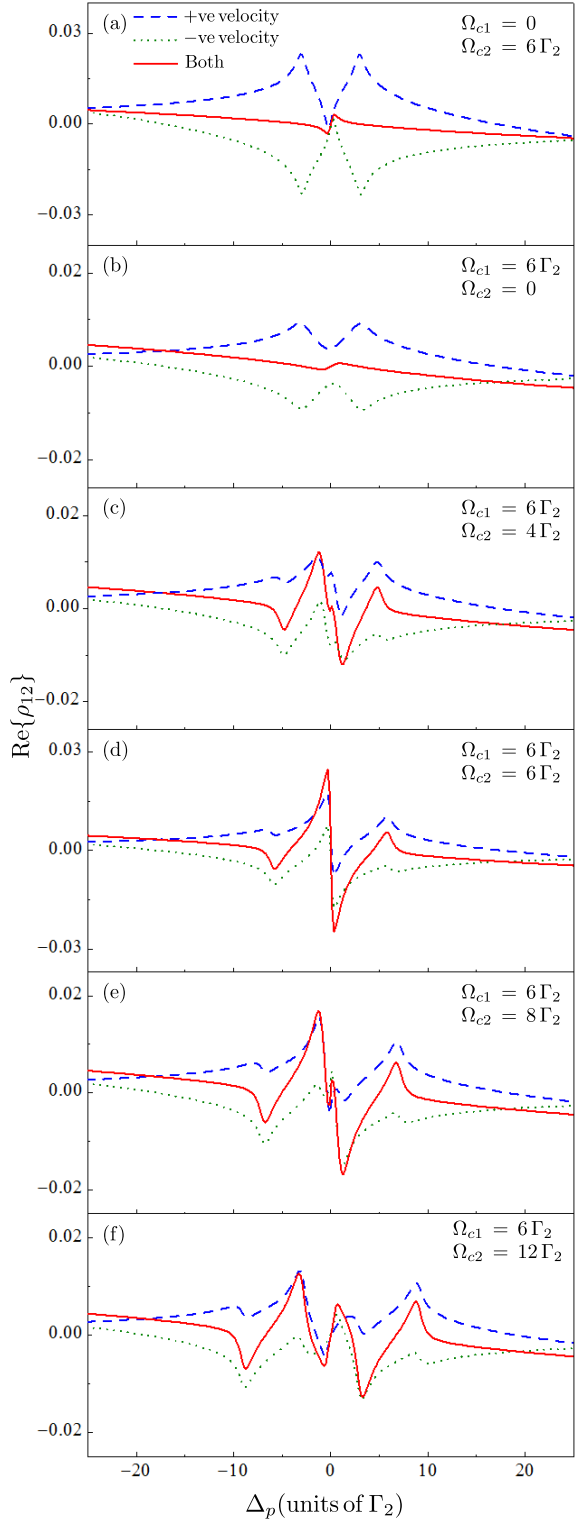}
\caption{(Color online) Effect of Doppler averaging on probe \textbf{dispersion} at room temperature. (a) $\Omega_{c1} = 0$, $\Omega_{c2} = 6 \, \Gamma_{2}$; (b) $\Omega_{c1} = 6 \, \Gamma_{2}$, $\Omega_{c2} = 0$; (c) $\Omega_{c1} = 6 \, \Gamma_{2}$, $\Omega_{c2} = 4 \, \Gamma_{2}$; (d) $\Omega_{c1} = 6 \, \Gamma_{2}$, $\Omega_{c2} = 6 \, \Gamma_{2}$; (e) $\Omega_{c1} = 6 \, \Gamma_{2}$, $\Omega_{c2} = 8 \, \Gamma_{2}$; (f) $\Omega_{c1} = 6 \, \Gamma_{2}$, $\Omega_{c2} = 12 \, \Gamma_{2}$. Solid curves are for all atoms, dashed curves are for atoms moving with positive velocities, and dotted curves are for atoms moving with negative velocities.}
\label{Fig.5}
\end{figure}

The same linewidth reduction is also seen for the dispersion profiles, shown in Fig.\ \ref{Fig.5}. As in the case of stationary atoms, the slope near line center changes from positive for EIT to negative for EIA, demonstrating the potential of switching the speed of light propagation.

\section{Conclusion}
In conclusion, we have investigated a four-level atomic system in vee + ladder configuration, with a weak probe field and two strong control fields. For application to a real system, we consider relevant low-lying energy levels in $^{87}$Rb. The coupled density-matrix equations are solved in steady state to calculate the absorption and dispersion profiles of the probe beam. The calculation is done for both stationary atoms and for atoms moving in room temperature vapor. In both cases, probe response shows an EIA window with negative dispersion at line center when the control fields have equal strengths. The induced absorption peak splits when the control fields are unequal, resulting in an EIT window with positive dispersion.  For moving atoms, Doppler averaging leads to significant linewidth narrowing for both absorption and dispersion profiles. These results can therefore be used for applications such as switching between sub- and super-luminal light propagation inside an atomic medium.

\section*{Acknowledgments}
This work was supported by the Department of Science and Technology, India. VB acknowledges financial support from a DS Kothari post-doctoral fellowship of the University Grants Commission, India.

\section*{Appendix}
In our analysis, the coupled density matrix equations in rotating frame are calculated by using the definition of density matrix and the expression for \emph{Hamiltonian}. The time evolution of system---by incorporating the decay of atoms from each level and repopulation from excited levels---is given by
\begin{eqnarray*}
\dot{\rho}_{11}&=&\Gamma_{2}\rho_{22}+\Gamma_{3}\rho_{33}+\dfrac{i}{2}\left(\Omega_{p}^{*}\rho_{12}-\Omega_{p}\rho_{21}\right)
\\&&+\dfrac{i}{2}\left(\Omega_{c1}^{*}\rho_{13}-\Omega_{c1}\rho_{31}\right);\nonumber\\
\dot{\rho}_{22}&=&-\Gamma_{2}\rho_{22}+\Gamma_{4}\rho_{44}+\dfrac{i}{2}\left(\Omega_{p}\rho_{21}-\Omega_{p}^{*}\rho_{12}\right)
\\&&+\dfrac{i}{2}\left(\Omega_{c2}^{*}\rho_{24}-\Omega_{c2}\rho_{42}\right);\nonumber\\
\dot{\rho}_{33}&=&-\Gamma_{3}\rho_{33}+\dfrac{i}{2}\left(\Omega_{c1}\rho_{31}-\Omega_{c1}^{*}\rho_{13}\right);\nonumber\\
\dot{\rho}_{44}&=&-\Gamma_{4}\rho_{44}+\dfrac{i}{2}\left(\Omega_{c2}\rho_{42}-\Omega_{c2}^{*}\rho_{24}\right);\nonumber\\
\dot{\rho}_{12}&=&\left(-\dfrac{\Gamma_{2}}{2}+i\Delta_{p}\right)\rho_{12}+\dfrac{i}{2}\Omega_{p}\left(\rho_{11}-\rho_{22}\right)
\\&&-\dfrac{i}{2}\left(\Omega_{c1}\rho_{32}-\Omega_{c2}^{*}\rho_{14}\right);\nonumber\\
\dot{\rho}_{13}&=&\left(-\dfrac{\Gamma_{3}}{2}+i\Delta_{c1}\right)\rho_{13}
+\dfrac{i}{2}\Omega_{c1}\left(\rho_{11}-\rho_{33}\right)
\\&&-\dfrac{i}{2}\Omega_{p}\rho_{23};\nonumber\\
\dot{\rho}_{14}&=&\left(-\dfrac{\Gamma_{4}}{2}+i(\Delta_{p}+\Delta_{c2})\right)\rho_{14}
\\&&+\dfrac{i}{2}\left(\Omega_{c2}\rho_{12}-\Omega_{p}\rho_{24}-\Omega_{c1}\rho_{34}\right);\nonumber\\
\dot{\rho}_{23}&=&\left(-\dfrac{\Gamma_{2}+\Gamma_{3}}{2}+i(\Delta_{c1}-\Delta_{p})\right)\rho_{23}
\\&&+\dfrac{i}{2}\left(\Omega_{c1}\rho_{21}-\Omega_{p}^{*}\rho_{13}-\Omega_{c2}\rho_{43}\right);\nonumber\\
\dot{\rho}_{24}&=&\left(-\dfrac{\Gamma_{2}+\Gamma_{4}}{2}+i\Delta_{c2}\right)\rho_{24}
+\dfrac{i}{2}\Omega_{c2}\left(\rho_{22}-\rho_{44}\right)
\\&&-\dfrac{i}{2}\Omega_{p}^{*}\rho_{14};\nonumber\\
\dot{\rho}_{34}&=&\left(-\dfrac{\Gamma_{3}+\Gamma_{4}}{2}+i(\Delta_{p}+\Delta_{c2}-\Delta_{c1})\right)\rho_{34}
\\&&+\dfrac{i}{2}\left(\Omega_{c2}\rho_{32}-\Omega_{c1}^{*}\rho_{14}\right);\nonumber\\
\end{eqnarray*}

The positive decay terms in $\dot{\rho_{ii}}$ ($i=1,2,3,4$) equations are
due to the fact that there is an increase in population of lower levels due to the decay of atoms from higher levels.

\section*{References}
%

\end{document}